\documentclass[a4paper]{spie}  

\usepackage{amsmath,amsfonts,amssymb}
\usepackage{graphicx}
\usepackage[colorlinks=true, allcolors=blue]{hyperref}
\usepackage{float}

\title{Design and implementation of an SLM-driven coherent
differential imaging scheme}

\author[a]{L. Lin}
\author[a]{L. Marquis}
\author[a]{R. Tandon}
\author[a]{J. Kühn}

\affil[a]{Division of Space and Planetary Sciences, University of Bern, Sidlerstrasse 5, 3012 Bern, Switzerland}

\authorinfo{Further author information: (Send correspondence to L.Lin)\\L.Lin: E-mail: liurong.lin@unibe.ch, Telephone: +41 31 684 85 76}

\begin{document} 
\maketitle

\begin{abstract}

Direct imaging of exoplanets from the ground remains fundamentally limited by fast-evolving post-adaptive optics (AO) residual wavefront errors, and by quasi-static speckles arising from non-common path aberrations (NCPAs) downstream of the AO wavefront sensor. To first order, both sources of contrast degradation remain coherent with the stellar light, and recent studies have shown that coherent differential imaging (CDI) techniques can address their quasi-static and slowly evolving components. Recent laboratory demonstrations using focal plane phase diversity with an active LCoS spatial light modulator (SLM) have further established that local phase modulation can distinguish coherent speckles from incoherent astrophysical signals. Here we present a framework for the design and implementation of an SLM-based CDI integrator. Through simulations, we study the effects of varying the number of phase steps, modulation ring width, planet separation, and planet-to-star contrast. The numerical results are further supported by an initial laboratory validation. In the future, with the emergence of ultra-low-noise fast near-infrared detectors such as avalanche photodiode and microwave kinetic inductance detector (MKID) arrays, time-domain focal-plane CDI detection schemes could potentially also be able to suppress rapidly varying post-AO residual wavefront errors.

\end{abstract}


\section{INTRODUCTION}
\label{sec:intro}
Ground-based direct imaging of exoplanets is limited by residual starlight that survives AO correction and coronagraphy. This residual light forms speckles that can hide or mimic faint planets. Post-AO atmospheric residuals create fast-evolving speckles that decorrelate on timescales from milliseconds to seconds, depending on seeing, wind speed, AO performance, and focal-plane location \cite{spekle_decorrelate}, while others are quasi-static and arise from non-common-path aberrations (NCPAs) downstream of the AO wavefront sensor \cite{Speckle_lifetime}. Since these speckles remain coherent with the stellar light to first order, they can be distinguished from incoherent astrophysical signals such as planets or disks using coherent differential imaging (CDI) \cite{CDI_first_paper}.

The idea behind coherent differential imaging (CDI) was first proposed using synchronous interferometric modulation to distinguish coherent stellar speckles from incoherent companion light \cite{CDI_first_paper}. Later, CDI was demonstrated on sky by combining a coronagraph with phase-shifting interferometry at Palomar \cite{CDI_bottom}, showing that coherence information could be used both to suppress speckles and to identify a real companion. More recent work extended CDI toward faster ground-based operation using self-coherent camera concepts, and CDI has also been demonstrated on VLT/SPHERE for the recovery of extended circumstellar disks while avoiding the self-subtraction artifacts commonly associated with ADI \cite{Potier_2025}.

Programmable focal-plane phase masks provide another route toward synchronous CDI. In particular, an active LCoS SLM can be used both as a coronagraphic focal-plane mask and as a temporal phase-modulation element, allowing selected components of the coherent stellar field to be modulated at a known frequency \cite{Jonas_CDI, Arikan2018}. In this work, we develop a focal-plane CDI framework using an active liquid-crystal-on-silicon spatial light modulator (LCoS SLM). By applying annular phase modulation in the focal plane, the SLM gives coherent stellar speckles a controlled temporal signature, while incoherent planet light remains largely unmodulated. We use simulations to study how CDI performance depends on the number of phase steps, modulation ring width, planet separation, and planet-star contrast. The current objective is to develop a robust SLM-based CDI instrument capable of suppressing quasi-static NCPA speckles and slowly evolving post-AO residuals. A future extension will target fast CDI operation, with the full modulation sequence completed within the atmospheric coherence time to probe rapidly varying post-AO residuals. This development is enabled by the emergence of fast, low-noise near-infrared detectors, including avalanche-photodiode and MKID arrays. The next step is to implement the method on a laboratory testbed using a 400 Hz SLM and a C-RED One detector, with the longer-term goal of integrating the technique into a future upgrade of the PLACID instrument on the 4-m DAG telescope.

\section{METHODS}
\label{sec:method}
\begin{figure}[H]
    \centering
    \includegraphics[width=0.8\linewidth]{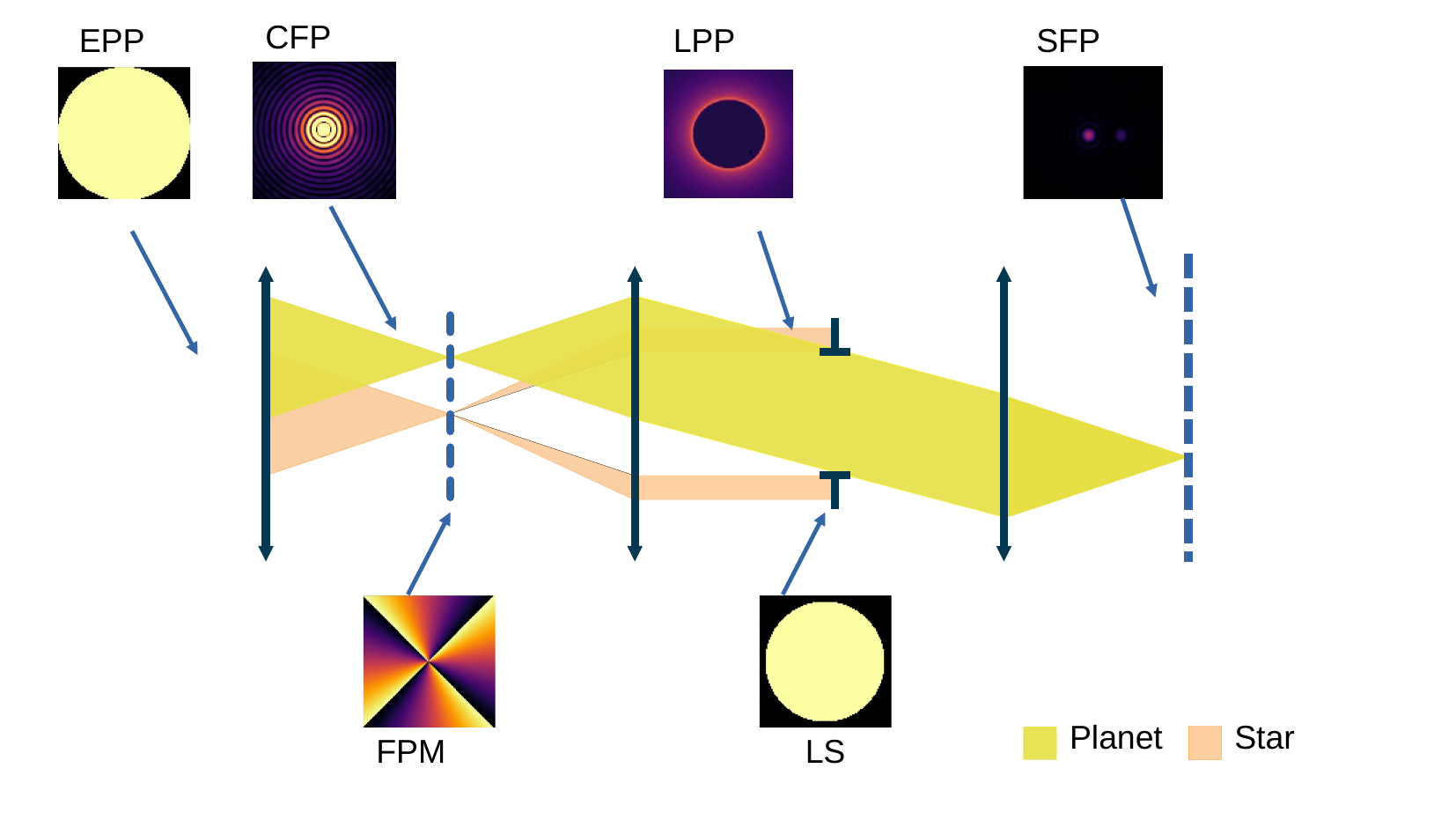}
    \caption{Cartoon illustration of light propagation through a coronagraphic system used for focal-plane CDI. Key optical planes are labeled, including the entrance pupil plane (EPP), coronagraphic focal plane (CFP), Lyot pupil plane (LPP), and science focal plane (SFP). The stellar and planetary light propagate differently through the coronagraph: the on-axis stellar light is diffracted by the focal-plane mask (FPM) and largely rejected by the Lyot stop (LS), while the off-axis planet signal passes through with much weaker attenuation. In the CDI scheme considered here, the FPM also serves as an active phase-modulation element, introducing controlled temporal phase offsets in the focal plane.}
    \label{fig:coro_scheme}
\end{figure}
\begin{figure}[H]
    \centering
    \includegraphics[width=1\linewidth]{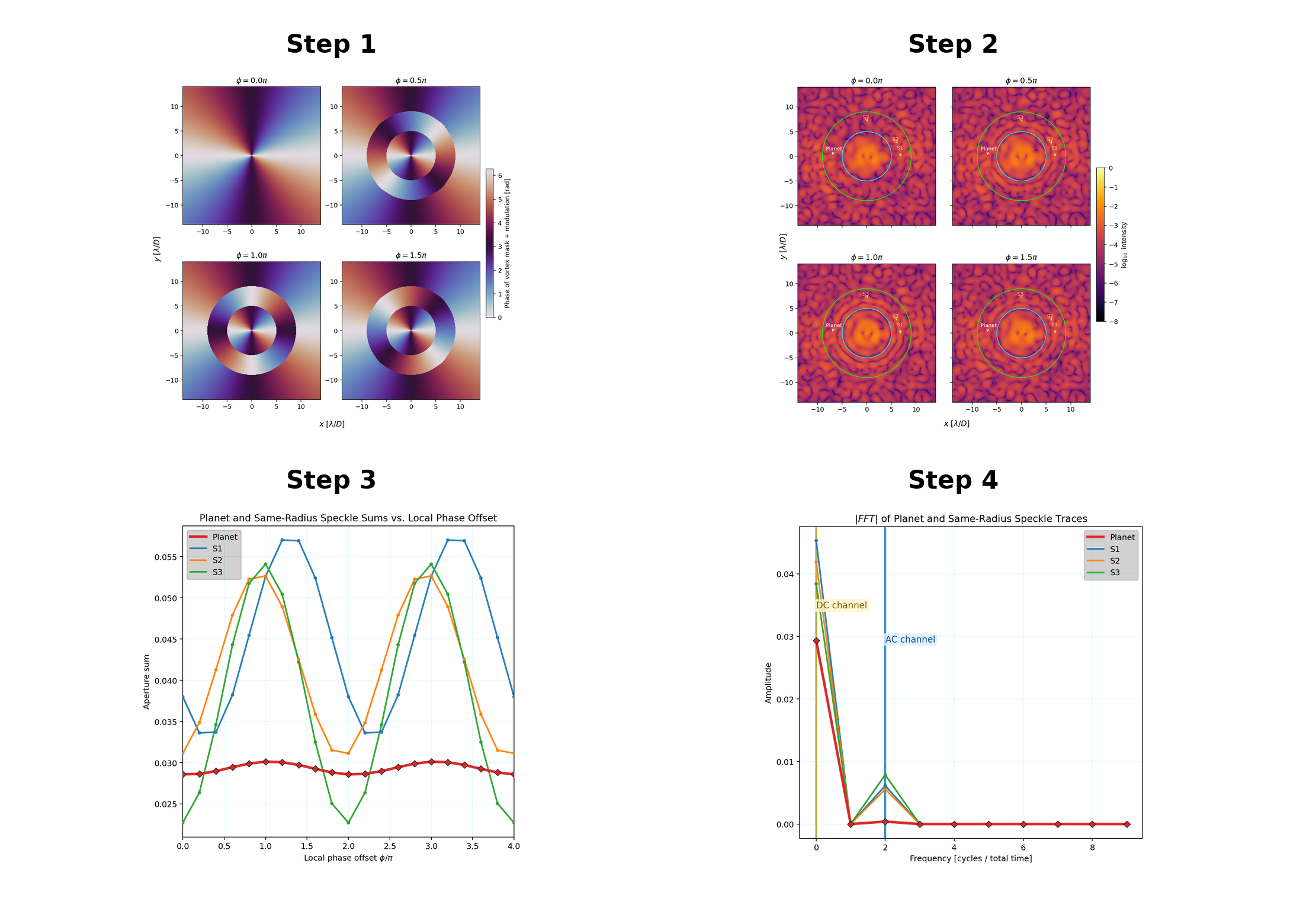}
    \caption{Workflow of the focal-plane phase-modulation coherent differential imaging method. Step 1: a prescribed local phase modulation is applied to the vortex coronagraph mask and stepped through a sequence of phase offsets. Step 2: for each phase step, a coronagraphic PSF is generated, producing a time series of focal-plane intensity images in which coherent stellar speckles respond to the modulation while the incoherent companion signal remains approximately stationary. Step 3: the figure shows an example intensity trace obtained from the aperture-summed signal at a selected location in the PSF as a function of modulation phase. Step 4: the figure shows the Fourier spectrum of that same aperture-summed temporal signal, where the DC term represents the stationary component and the nonzero modulation-frequency terms represent the coherent phase-driven response. In the actual simulation, this analysis is not performed on a single aperture-summed trace. Instead, the temporal FFT is computed independently at every focal-plane pixel of the full PSF time-series cube, and the resulting pixel-wise DC and modulation-frequency (AC) amplitudes are used to construct coherence and incoherence maps.}
    \label{fig:workflow}
\end{figure}

Before describing the numerical CDI workflow, we first introduce the optical layout considered in this work. Figure~\ref{fig:coro_scheme} shows a schematic illustration of light propagation through the coronagraphic system, including the entrance pupil plane (EPP), coronagraphic focal plane (CFP), Lyot pupil plane (LPP), and science focal plane (SFP). In this configuration, the focal-plane mask (FPM) plays a dual role: it provides the coronagraphic phase mask used to suppress the on-axis stellar light, and it also serves as the active phase-modulation element used for CDI. The workflow described below builds on this optical configuration by applying a controlled temporal phase offset at the FPM and analyzing the resulting focal-plane intensity response.

The analysis procedure is summarized in the workflow image (Fig. \ref{fig:workflow}), which illustrates the four main stages of the simulation and signal-separation pipeline.
In Step 1 of Fig. \ref{fig:workflow}, a localized focal-plane phase modulation is applied to the coronagraphic system. In the present simulations, this modulation is imposed on top of the vortex phase mask and stepped through a prescribed sequence of phase offsets spanning one or more modulation cycles. The purpose of this step is to introduce a controlled temporal perturbation that affects the coherent stellar field while leaving the incoherent companion contribution unchanged in expectation. We use 10 phase steps per modulation cycle in the simulations presented here.

In Step 2 of Fig. \ref{fig:workflow}, a coronagraphic PSF is generated for each phase step, producing a time-ordered cube of SFP images. Because the stellar speckle field is coherent with the stellar light, the modulated and unmodulated field components interfere, causing the speckle intensities to vary systematically from frame to frame. By contrast, the incoherent companion signal does not participate in the same interference process and therefore remains approximately stationary across the modulation sequence.

The temporal signatures of the coherent and incoherent components are illustrated schematically in Step 3 of Fig. \ref{fig:workflow}. This panel shows example aperture-summed intensity traces as a function of modulation phase at selected locations in the PSF. The purpose of the aperture-summed intensity is purely illustrative: it highlights that an incoherent companion contribution is expected to vary only weakly over the modulation cycle, whereas coherent speckle signals exhibit a stronger periodic response. These aperture-summed traces are not, however, the primary analysis product used in the simulation; they are included only to provide an intuitive picture of the temporal behavior that underlies the frequency-domain separation.

The quantitative analysis is summarized in Step 4 and is carried out on the PSF time-series cube. For each focal-plane pixel $(x,y)$, we construct the intensity sequence $I(x,y,\phi_k)$ over the discrete phase steps $\phi_k$ and compute its discrete Fourier transform along the modulation axis. In practice, the DC term is used as a proxy for the stationary contribution, while the selected modulation-frequency term is used as a proxy for the coherent response. Applying this FFT independently to every pixel in the focal plane yields spatial maps of the static and modulated components. Coherence and incoherence metrics are then constructed from these pixel-wise Fourier amplitudes. The relative amplitudes of the DC and AC terms provide a spatial criterion for distinguishing regions dominated by coherent stellar speckles from regions containing a stronger incoherent companion contribution. Let \(I(x,y;\phi)\) denote the intensity at pixel \((x,y)\) for applied phase offset \(\phi\).  
Its discrete Fourier transform along the phase axis is written as
\[
\tilde{I}(x,y;f) = \mathcal{F}_{\phi}\{I(x,y;\phi)\}(f),
\]
where \(f=0\) denotes the DC component and \(f_{\mathrm{mod}}\) is the selected modulation frequency.

The coherence map is then defined as
\[
C(x,y) = \frac{\left|\tilde{I}(x,y;f_{\mathrm{mod}})\right|}{\left|\tilde{I}(x,y;0)\right|}.
\]

The incoherence map is defined as the inverse of the coherence map,
\[
I_{\mathrm{incoh}}(x,y) = \frac{1}{C(x,y)}
= \frac{\left|\tilde{I}(x,y;0)\right|}{\left|\tilde{I}(x,y;f_{\mathrm{mod}})\right|}.
\]

\subsection{SNR calculation}

The detectability of the planetary signal is quantified using an aperture-based signal-to-noise ratio (SNR) measured on the incoherence map. A circular aperture of radius 1 $\lambda/D$ is centered on the planet location, and the mean map value inside this aperture is defined as the planet signal, \(S_{\mathrm{planet}}\). Background is estimated from a set of identical apertures placed along an annulus at the same angular separation from the star as the planet location. If \(S_k\) denotes the mean value in the \(k\)-th background aperture, the background mean and standard deviation are
\[
\mu_{\mathrm{bg}} = \frac{1}{N_{\mathrm{bg}}}\sum_{k=1}^{N_{\mathrm{bg}}} S_k,
\qquad
\sigma_{\mathrm{bg}} =
\sqrt{
\frac{1}{N_{\mathrm{bg}}}
\sum_{k=1}^{N_{\mathrm{bg}}}
\left(S_k - \mu_{\mathrm{bg}}\right)^2
}.
\]

For the incoherence map, the SNR is defined as
\[
\mathrm{SNR} = \frac{S_{\mathrm{planet}} - \mu_{\mathrm{bg}}}{\sigma_{\mathrm{bg}}},
\]
where a larger value indicates that the planet aperture is brighter than the local background. 

\subsection{Laboratory validation}
\label{sec:method_lab}

To validate the simulation results, we performed laboratory measurements on an optical testbed. The stellar and planetary sources were simulated using two monochromatic laser beams operating at 1550~nm. The planet-to-star intensity ratio was set to $2\times10^{-3}$. A phase screen with 160~nm RMS wavefront error was introduced to generate a controlled speckle field. The entrance pupil was defined by a circular aperture without central obstruction. Both the charge-2 vortex focal-plane phase mask and the temporal local phase modulation were implemented with a $1024 \times 1024$ Meadowlark Optics Ultra-High Power spatial light modulator. The resulting coronagraphic images were recorded using a First Light Imaging C-RED 2 detector. For each modulation cycle, ten phase steps were applied. Three modulation cycles were recorded and combined to generate each coherence and incoherence map.

\section{RESULTS}
\label{sec:result}
\begin{figure}[H]
    \centering
    \includegraphics[width=.9\linewidth]{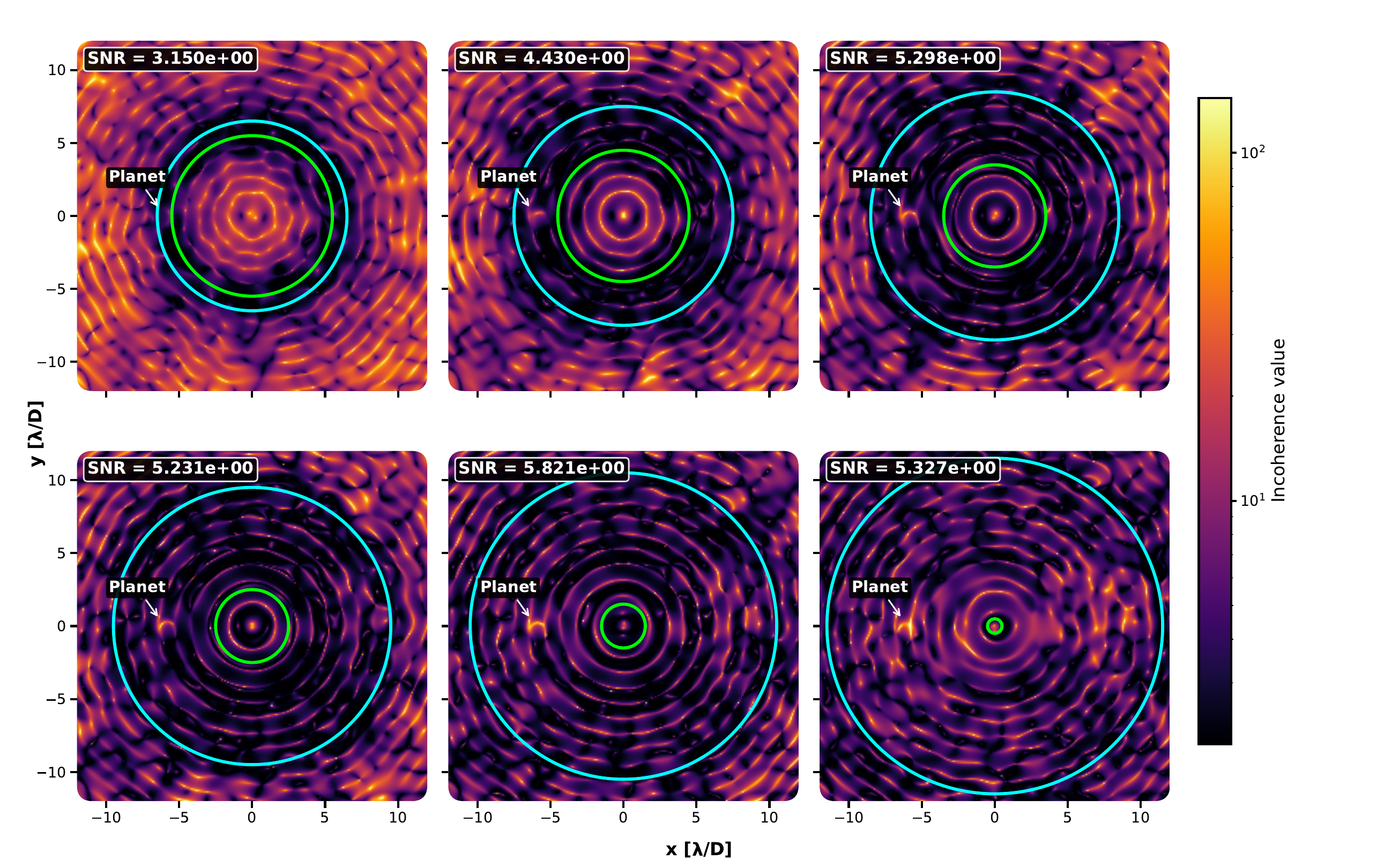}
    \caption{Incoherence maps for a planet located at $r = 6\,\lambda/D$ and $\theta = 180^\circ$. Each panel corresponds to a different ring-shaped phase-modulation region of interest (ROI), with ring widths ranging from $w = 1\,\lambda/D$ to $w = 11\,\lambda/D$ in steps of $2\,\lambda/D$. The planet position is marked, and the corresponding recovered SNR is indicated above each panel. The results were obtained using a charge-2 vortex focal-plane mask and two modulation cycles, each consisting of 10 phase steps.}
    \label{fig:incoherence}
\end{figure}

Figure \ref{fig:incoherence} shows that, as the size of the ring-shaped ROI increases, the recovered SNR in the incoherence maps rises from approximately 3.15 for the smallest ROI (ring width w = 1 $\lambda$/D) to 5.82 for intermediate-to-large ROIs, before decreasing slightly to about 5.31 for the largest ROI (ring width w = 11 $\lambda$/D). The highest SNR is obtained for $w=9\,\lambda/D$, while both $w=7\,\lambda/D$ and $w=9\,\lambda/D$ provide a relatively clear separation of the planet from the surrounding residual field.
The CDI performance also degrades as the angular separation between the star and planet is reduced, because the companion becomes increasingly difficult to distinguish from the brighter and more strongly structured stellar residuals at small separations. Increasing the number of phase steps beyond 10 provides only a limited additional improvement, indicating that a 10-step modulation sequence already samples the phase-dependent intensity response sufficiently well for the configurations considered here.

\begin{figure}[H]
    \centering
    \includegraphics[width=.9\linewidth]{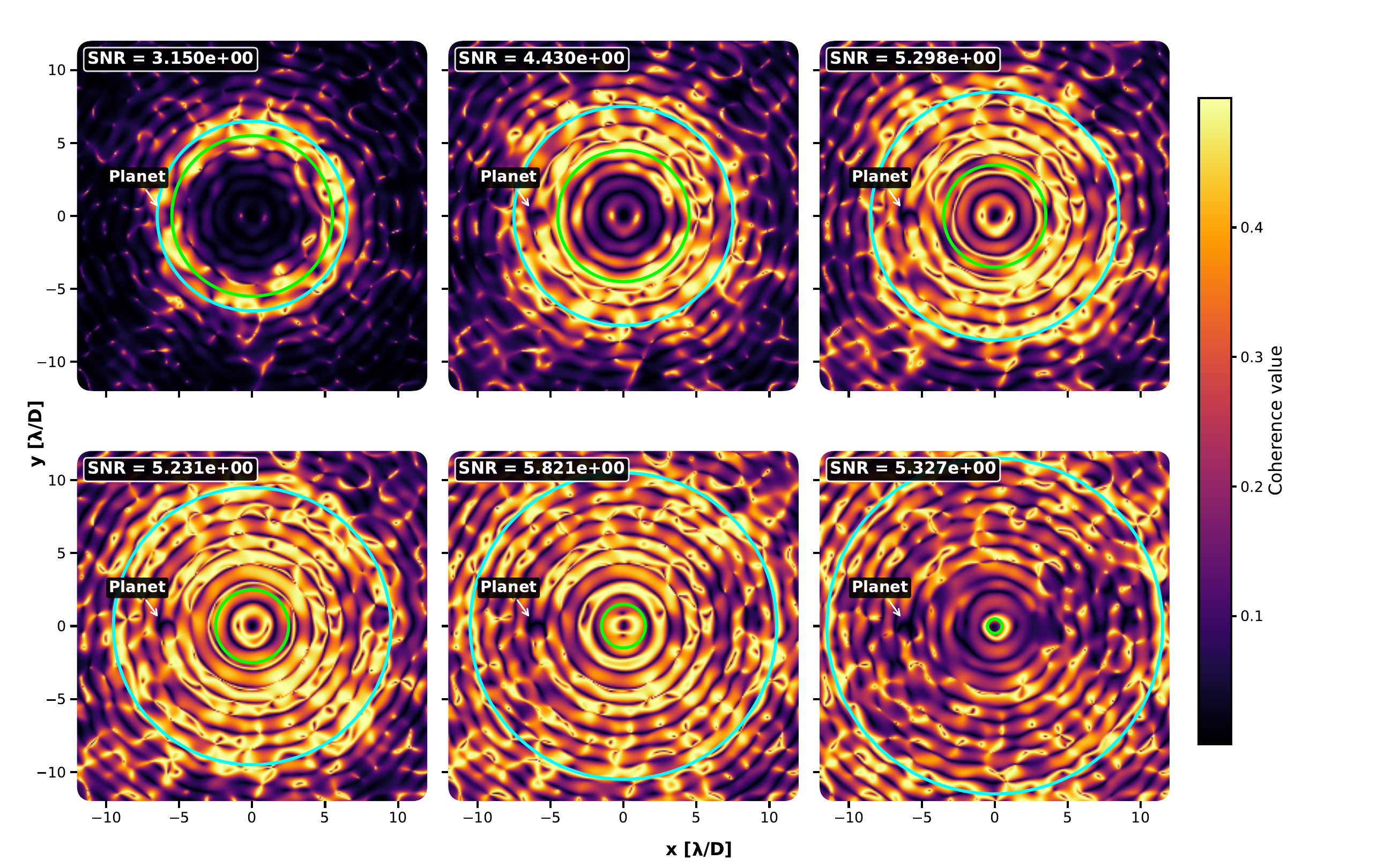}
    \caption{Coherence maps for a planet located at $r = 6\,\lambda/D$ and $\theta = 180^\circ$. Each panel corresponds to a different ring-shaped phase-modulation region of interest (ROI), with ring widths ranging from $w = 1\,\lambda/D$ to $w = 11\,\lambda/D$ in step of $2\,\lambda/D$. The planet position is marked, and the corresponding incoherence map SNR is indicated above each panel. The results were obtained using a charge-2 vortex FPM and two modulation cycles, each consisting of 10 phase steps.}
    \label{fig:coherence}
\end{figure}

The sinusoidal variation of the stellar speckles arises from interference between the phase-modulated and unmodulated parts of the coherent stellar field. At a given image location, the field can be written as
\[
E_p(\phi) = E_{u,p} + E_{m,p} e^{-i\phi},
\]
where \(E_{u,p}\) is the unmodulated stellar contribution and \(E_{m,p}\) is the modulated stellar contribution. The corresponding intensity is
\[
I_p(\phi) = |E_{u,p}|^2 + |E_{m,p}|^2 + 2\,\Re\!\left\{E_{u,p} E_{m,p}^* e^{i\phi}\right\},
\]
so the last term produces the phase-dependent sinusoidal modulation. This is the key reason why stellar speckles, being coherent with the parent stellar light, exhibit a strong modulation signal in the CDI analysis.

By contrast, the planetary signal is incoherent with the stellar field and therefore does not produce the same phase-locked interference term with the modulated stellar light. This difference is what enables CDI to distinguish coherent stellar speckles from incoherent companion light.

When the phase-modulated annulus is enlarged, the recovered SNR does not necessarily increase because the modulation strength is governed primarily by the balance between the unmodulated and modulated field components, $E_{u,p}$ and $E_{m,p}$, at each science-image pixel. Although a wider annulus modulates a larger fraction of the stellar field, this does not guarantee a stronger interference signal. If the annulus becomes too large, the relative amplitudes of $E_{u,p}$ and $E_{m,p}$ may become less favorable, causing the effective cross term
\[
E_{u,p}E_{m,p}^{*}
\]
to decrease in magnitude. In addition, $E_{m,p}$ is formed by the coherent sum of contributions propagated from the entire modulated region, which may combine with different complex phases and partially cancel. The strongest sinusoidal response, and therefore the highest SNR, is thus obtained for an intermediate ROI size that provides a favorable balance between $E_{u,p}$ and $E_{m,p}$, rather than simply maximizing the amount of modulated stellar light.

\subsection{Laboratory result} 
\begin{figure}[H]
    \centering
    \includegraphics[width=.9\linewidth]{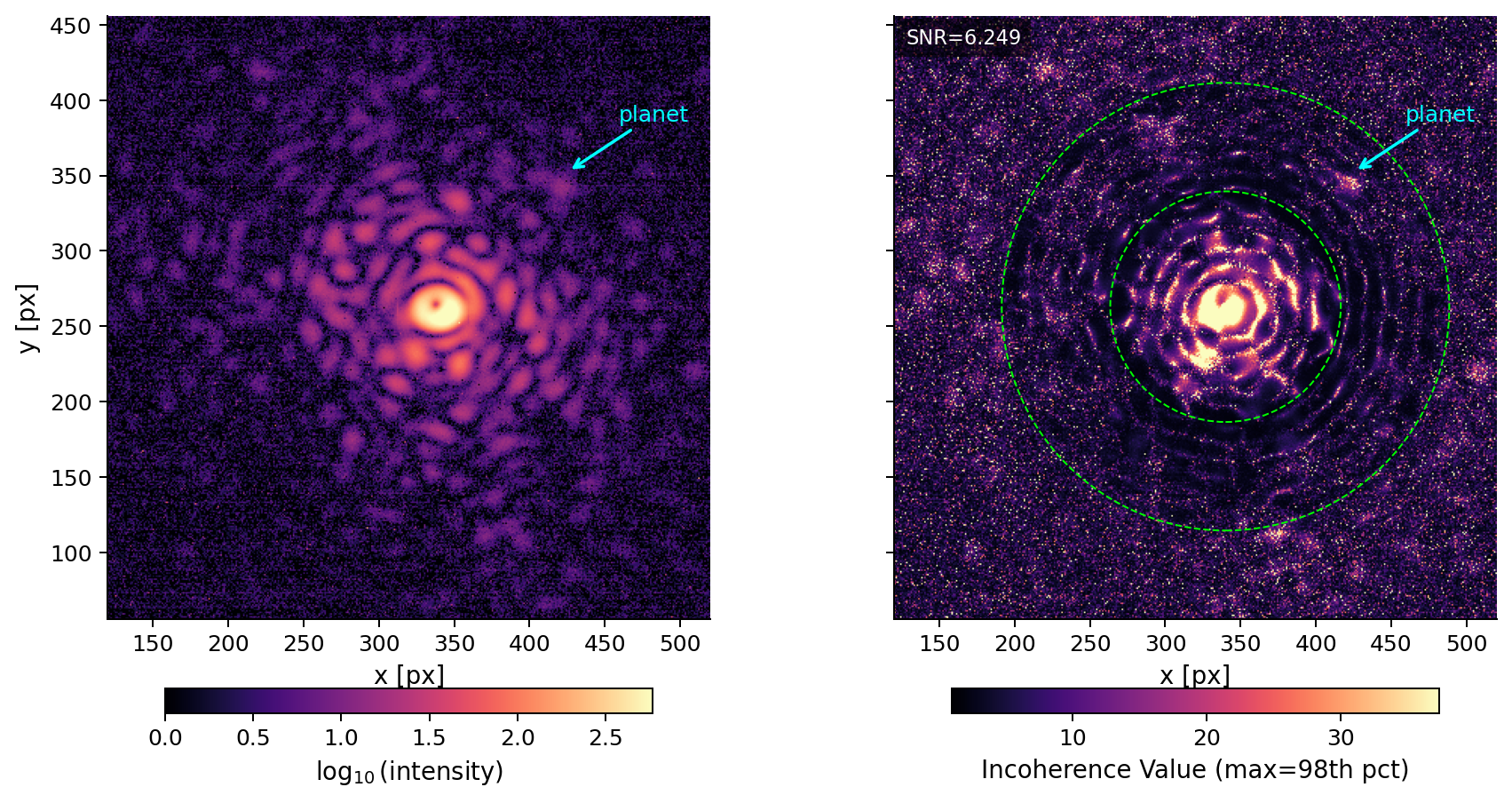}
    \caption{Laboratory measurement of the raw science PSF and incoherence map with the phase-modulation ring of w = 6 $\lambda$/D. Left: log-scaled PSF intensity, with the planet location marked. Right: corresponding incoherence map, where the annular ring denotes the applied phase modulation region and the detected source is identified with an estimated SNR of 6.249.}
    \label{fig:lab}
\end{figure}

Figure \ref{fig:lab} shows the raw PSF and the corresponding incoherence map measured in the laboratory using a phase-modulation ring with $w=6\,\lambda/D$. The left panel shows the log-scaled raw SFP, in which the planet signal is embedded within the stellar residuals and surrounding speckle field. The right panel shows the corresponding incoherence map, where the dashed annulus marks the phase-modulation region and the planet appears as a localized bright feature outside the central core. The SNR of the detected planet signal is 6.249, indicating improved visibility in the incoherence map relative to the raw intensity image.

\section{CONCLUSIONS and OUTLOOK}
\label{sec:con}
We have presented a CDI framework based on localized temporal phase modulation applied directly with a SLM as an FPM. In this approach, the coronagraphic FPM serves both as the stellar-suppression element and as the phase-modulation device. The imposed phase sequence gives coherent stellar speckles a controlled temporal signature, whereas an incoherent planetary signal remains primarily in the stationary component. Pixel-wise Fourier-domain post processing of the resulting image cube can therefore be used to construct coherence and incoherence maps and to distinguish an off-axis companion from the surrounding speckle field.

The simulations demonstrate that the size of the phase-modulated annulus is an important parameter because it determines the relative amplitudes of the modulated and unmodulated components of the coherent stellar field. The temporal modulation signal arises from interference between these two components and is therefore strongest when both contribute significantly at the science-image given location. If the modulated fraction is too small, the phase-dependent response remains weak. Conversely, if too much of the stellar field is modulated, the remaining unmodulated reference contribution becomes insufficient, which can also reduce the interference term. In addition, the optical coupling of different parts of the modulated region to a given science-image pixel may involve different complex phases, leading to constructive or destructive interference. The optimal annular geometry is therefore set by the balance between the modulated and unmodulated stellar contributions, rather than by the size of the modulated area alone.

The laboratory measurements provide an initial experimental validation of the method. Using a charge-2 vortex implemented on an LCoS SLM, ten phase steps per cycle and three modulation cycles, the incoherence reconstruction recovered the injected companion at 10 $\lambda/D$ with an estimated SNR of 6.249. This result confirms that the temporally modulated stellar contribution can be separated from an incoherent source under controlled laboratory conditions, despite the presence of a 160-nm RMS aberrated wavefront (central wavelength = 1550 nm).

However, the current method should presently be regarded primarily as a planet-detection or planet-finding technique. Because residual speckles can partially overlap the companion signal, the spatial shape and photometric distribution of the recovered source in the incoherence map may be distorted by the local coherent speckle field. The reconstructed feature therefore provides evidence for the presence and approximate location of an incoherent source, but should not necessarily be interpreted as an accurate representation of its intrinsic point-spread function or morphology.

Several developments are required before the technique can be applied under realistic observing conditions. First, a broader parameter study will investigate the joint dependence on modulation  geometry, number of phase steps, number of cycles, planet separation and planet-to-star contrast. The modulation sequence should also be optimized by considering detector noise, photon noise, SLM response time and the total integration time required to complete one CDI cycle. These factors will determine the minimum companion brightness that can be recovered and the maximum speckle-evolution rate that can be followed.

Future simulations and laboratory experiments will additionally include broadband illumination. Broadband operation is particularly important because both the coronagraphic response and the spatial scale of the modulated region vary with wavelength. The robustness of the Fourier-domain separation should therefore be evaluated over realistic near-infrared bandwidths.

The next experimental step will be to implement the CDI sequence with a fast 400-Hz SLM and a low-noise C-RED One detector. This configuration is intended to shorten the modulation cycle sufficiently to probe speckles evolving on timescales comparable to, or shorter than, the atmospheric coherence time. In the longer term, the method is planned as a possible operational mode for the PLACID instrument on the 4-m DAG telescope. On-sky testing will determine whether fast focal-plane phase modulation can extend CDI beyond quasi-static speckle calibration and provide real-time discrimination between rapidly evolving coherent residual starlight and incoherent astrophysical sources such as exoplanets and circumstellar disks.

\section{ACKNOWLEDGEMENTS}
\label{sec:ac}
The RACE-GO project has received funding from the Swiss State Secretariat for Education, Research and Innovation 
(SERI), under the ERC replacement scheme following the discontinued participation of Switzerland to Horizon Europe. 
Part of this work has been carried out within the framework of the National Centre of Competence in Research PlanetS 
supported by the Swiss National Science Foundation under grants 51NF40 182901 and 51NF40 205606. 

The authors used ChatGPT (OpenAI) to assist with English grammar and spelling, and clarity editing during manuscript preparation. Example prompt: “Please correct this paragraph for grammar without changing the original meaning.” All suggested edits were reviewed and approved by the authors, who take full responsibility for the final text.

\newpage
\bibliography{report} 
\bibliographystyle{spiebib} 

\end{document}